# Attribution Bias in Philosophical Knowledge Graphs: Corpus Frequency versus Temporal Sourcing

Joy Bose
Independent Researcher
joy.bose@ieee.org

**Abstract**

Computational knowledge graphs typically assign philosophical concepts to traditions based on corpus frequency: the school that mentions a concept most becomes its attributed tradition. We argue this conflates three distinct measurements: textual power, historical priority, and philosophical significance, and demonstrate the consequences using the darshana-graph, a knowledge graph of 28,322 documented relationships across Hindu, Buddhist, and Jain traditions. Seven of the top 25 concepts by betweenness centrality predate their attributed school by 288 to 2,288 years. Moksha, attributed to Advaita Vedanta, appears first in Jain sources over 1,200 years earlier. The most reliable snapshot of the network, at 300 BCE using only explicitly dated sources, shows a genuinely pluralistic structure: 59% Vedic, 24% Jain, 18% Buddhist, not the Advaita-dominated picture that corpus frequency produces. We also quantify a critical distortion in the temporal method itself: between 300 CE and 800 CE the network grows from 18 to 1,028 nodes, with 97.4% carrying Advaita proxy dates, revealing that apparent Advaita dominance reflects textual survival, not philosophical history. Beyond correcting attribution bias, the temporally grounded graph enables a new class of analysis: structural homology across traditions. Applying ego-network feature vectors to 48 temporally labelled concepts across eight traditions, we identify cross-tradition concept pairs with high structural similarity. The method recovers known scholarly correspondences, including purusha-jiva (Samkhya/Jain, sim 0.990) and prakriti-maya (Samkhya/Vedic, sim 0.972), and surfaces novel structural homologies. Nibbana (Buddhist liberation) and samsara (Vedic cyclic existence) score 0.954 despite being doctrinal opposites, because both function as the ultimate reference concept in their tradition's soteriology. Cetana (Buddhist intention) and ajiva (Jain non-living matter) score 0.923, a pairing absent from the scholarly literature that the network identifies as structurally analogous. These results are not claims of doctrinal equivalence but of measurable structural homology: different philosophical vocabularies navigating a shared conceptual space. Corpus-frequency attribution, by assigning all high-betweenness concepts to Hindu schools, makes this shared structure invisible. Temporal attribution begins to restore it.

## 1. Introduction

Computational knowledge graphs have become a significant tool in digital humanities for mapping the structure of ideas across traditions, periods, and texts. A typical philosophical knowledge graph represents concepts as nodes and documented relationships between concepts as edges. Such graphs have been used to study intellectual influence, conceptual transmission, and the structural organisation of philosophical traditions in contexts ranging from Chinese classical thought to medieval European scholasticism to the history of science (Collins 1998; Moretti 2005; Karsdorp et al. 2021).

A central operation in building any such graph is attribution: assigning each concept to the tradition, school, or author most closely associated with it. This attribution shapes every downstream analysis. When a researcher asks which tradition has the most structurally central concepts, or which tradition acts as a bridge between other traditions, the answer depends entirely on which tradition each concept has been assigned to. Attribution is not a neutral preprocessing step. It is an interpretive act that encodes assumptions about philosophical history before any analysis begins.

In practice, most computational concept graphs assign each concept to the school or tradition that mentions it most frequently in the available corpus. We call this corpus-frequency attribution. It has a clear practical advantage: it requires no additional annotation beyond what is already in the dataset. But it encodes a hidden assumption that deserves to be stated explicitly: that frequency of textual use correlates with, or is a proxy for, philosophical priority. Under this assumption, the tradition that discussed a concept most extensively in surviving texts is treated, computationally, as the tradition most associated with that concept.

We argue that this assumption is problematic in a systematic and measurable way. It conflates three distinct things that are often confused in discussions of philosophical influence: textual power (which tradition produced and preserved the most text discussing a concept), historical priority (which tradition first documented the concept in a datable source), and philosophical significance (which tradition developed the most philosophically sophisticated or original treatment of the concept). These are three different questions, and the same computational method cannot answer all three. A tradition that produced more text, or whose texts are better preserved, or whose texts have been more extensively digitised, will appear to own more concepts regardless of whether it introduced those concepts, regardless of whether its treatment of them is the most philosophically important, and regardless of whether the attribution reflects historical reality.

This paper demonstrates the problem concretely and proposes a partial corrective, while being explicit about the limitations of that corrective.

We use the darshana-graph as our case study (Bose 2025). The darshana-graph (from the Sanskrit darshana, meaning philosophical view or school) is a knowledge graph of 28,322 documented relationships between philosophical concepts across Hindu, Buddhist, and Jain traditions, available publicly at joyboseroy/darshana-graph on HuggingFace. When we compute betweenness centrality using corpus-frequency attribution, every one of the top 25 concepts by betweenness is attributed to Hindu schools, with Advaita Vedanta dominating. When we apply temporal attribution using scholarly-cited source dates, the picture at 300 BCE is substantially different: the network, based on 17 concepts with explicit scholarly citations, is 59% Vedic, 24% Jain, and 18% Buddhist. Seven of the top 25 concepts by betweenness have earliest documented attestations that predate the school to which corpus-frequency attribution assigns them, with gaps ranging from 288 to 2,288 years. The concept moksha, attributed to Advaita Vedanta by corpus frequency, has its earliest available systematic soteriological attestation in Jain sources approximately 1,288 years before Advaita Vedanta existed as a school.

We also identify and quantify a critical distortion in the temporal method itself: between 300 CE and 800 CE, the network grows from 18 to 1,028 nodes, with 97.4% of new nodes carrying proxy dates from Advaita Vedanta's school founding date. This growth does not reflect philosophical innovation at that date. It reflects the scale of Advaita's commentarial output, the partial destruction of competing Buddhist manuscript traditions, and the digitisation advantage of Sanskrit texts in the available corpus. The 800 CE snapshot of the network is not a picture of Indian philosophy in 800 CE. It is a picture of what the surviving digitised corpus looks like at that date. This finding is not a failure of the method. It is the method making a corpus composition problem visible as a measurable quantity, which is itself a contribution.

Our argument is not that temporal sourcing is correct and corpus frequency is wrong. Our argument is that both methods are measuring something real and something limited, that these limitations are different in kind, and that understanding the difference is more informative than using either method in isolation. Corpus-frequency attribution measures textual power. Temporal attribution with scholarly citations measures earliest documented attestation. Neither measures philosophical significance. Conflating these three things produces systematically misleading results. Keeping them distinct is the methodological contribution of this paper. The measurement problem is not specific to Indian philosophy: it arises wherever computational knowledge graphs are built from corpora with uneven textual survival across traditions.

### 1.2 Measurement objectives: three questions, three methods

Before presenting our methodology, we introduce a framework that we believe should precede any analysis of a philosophical concept graph. We call it the measurement objectives framework.

A concept graph can be used to answer at least three distinct kinds of questions, and different questions require different attribution methods:

**Question 1: Which tradition discussed this concept most?** If this is the research question, corpus-frequency attribution is the appropriate method. It correctly answers this question within the limits of the available corpus. The answer tells us something real: which tradition had the resources, institutional continuity, and textual output to engage most extensively with a concept in the surviving literature.

**Question 2: Which tradition first documented this concept in a datable source?** If this is the research question, temporal attribution with scholarly-cited source dates is the appropriate method. It attempts to answer this question with explicit uncertainty ranges and citations. The answer is an earliest documented attestation, not a claim about origin in any deeper sense, and it is only as reliable as the scholarly dating literature it draws upon.

**Question 3: Which tradition developed the most philosophically sophisticated or original treatment of this concept?** If this is the research question, neither method is appropriate. This question requires philosophical argument, textual analysis, and historiographical judgment that no computational method can substitute for.

The problem we identify is that most computational concept graph papers are implicitly answering Question 1 while presenting their results as if they were answering Question 2 or Question 3. Making the measurement objective explicit before analysis begins is a minimal requirement for methodological honesty. We return to these three questions throughout the paper and use them as an organising framework for our discussion of results.

## 2. Background and Related Work

### 2.1 Network analysis in intellectual history

The application of network analysis to intellectual history has a substantial lineage in both sociology and digital humanities. Randall Collins's *The Sociology of Philosophies* (1998) established the foundational framework for treating philosophical traditions as social networks. Collins argued that intellectual change is driven not by isolated individual genius but by structured patterns of network rivalries and alliances, operating under what he calls the law of small numbers: the attention space of any active intellectual community is structurally constrained to support between three and six competing positions simultaneously. Crucially for the present paper, Collins showed that the success, expansion, or decline of intellectual coalitions is deeply contingent on shifts in their material bases, the institutional structures, patronage systems, monastic organizations, and preservation infrastructures that determine which networks survive and which contract. When material bases shift, opportunities open for certain networks to proliferate while others are forced to dissolve.

Stefan Jänicke and colleagues (2015) have catalogued visualisation approaches for digital humanities text analysis, noting both the analytical power of network representations and the risks of misreading visual spatial arrangements as semantic relationships. This concern connects directly to Johanna Drucker's (2011) foundational critique: digital humanities tools borrowed from the empirical sciences carry assumptions of knowledge as observer-independent and objective, which humanistic data cannot support. Drucker's distinction between data (presumed given) and capta (taken, constructed, selected) is central to our argument. The darshana-graph is capta: every edge it contains reflects a decision, about which texts to include, which relationships to extract, which attribution scheme to apply. Corpus-frequency attribution naturalises these decisions by making them invisible. Our temporal attribution layer attempts to make some of them visible and challengeable, while creating new and different forms of invisibility that we have tried to document explicitly.

### 2.2 Robustness of network analysis under archival decay

A specific and directly relevant body of literature concerns the robustness of network centrality metrics when applied to historical datasets that are inevitably incomplete. Yann Ryan and Sebastian Ahnert (2021), in their study of early modern correspondence networks published in the *Journal of Cultural Analytics*, conducted systematic robustness experiments by simulating archival decay on historical datasets. They modelled random document deletion, missing chronological blocks, and entity disambiguation errors, testing how network centrality rankings changed under different patterns of data loss. Their key finding is that betweenness centrality is moderately to highly robust under random archival decay. However, their model assumes random loss, which does not hold for the darshana-graph: the destruction of Buddhist monastic universities was systemic and targeted, precisely the regime where robustness guarantees break down.

### 2.3 The darshana-graph dataset

The darshana-graph (Bose 2025, arXiv:2606.18222) is a philosophical knowledge graph of 28,322 typed relationship edges across 2,349 concept nodes, extracted from a parallel commentary corpus covering Hindu, Buddhist, and Jain traditions. The dataset is available publicly at joyboseroy/darshana-graph on HuggingFace.

The full corpus underlying the darshana-graph contains approximately 125,040 text records. Of these, 114,591 are Pali Canon records, making the Pali Canon the numerically dominant source in the full corpus. However, the knowledge graph was extracted primarily from an 8,500-record parallel commentary subset that aligns primary root verses (from sources including the Brahma Sutras, the Bhagavad Gita, and the principal Upanishads) across eighteen commentators representing five schools of Vedanta and related traditions. This parallel commentary subset is structurally dominated by Sanskrit Vedantic commentators. The large numerical presence of the Pali Canon in the full corpus does not translate proportionally into the knowledge graph because the graph extraction pipeline was applied differently to the commentary subset than to the Pali Canon records.

Approximately 70% of extracted edges carry a general school attribution rather than a commentator-specific label, reflecting the limits of what the extraction pipeline could reliably attribute. The attribution bias analysis in this paper operates at the school level rather than the commentator level, and therefore inherits this 70% aggregate attribution rate without further amplification.

### 2.4 Corpus composition and the Sanskrit digitisation advantage

The compositional biases of the darshana-graph's source corpus are not unique to this dataset. They reflect structural features of the available Sanskrit digital archive that bear directly on any corpus-frequency analysis of Indian philosophical texts.

Sheldon Pollock's *The Language of the Gods in the World of Men* (2006) traces the rise of what he calls the Sanskrit cosmopolis: Sanskrit's transformation from a localised ritual language into a dominant vehicle of poetry, polity, and elite intellectual culture across South and Southeast Asia, sustained from approximately the beginning of the Common Era through the early modern period. This cosmopolitan status brought Sanskrit textual traditions continuous royal patronage, political elite support, and copying infrastructure unavailable to Prakrit, Pali, and vernacular traditions. The result is a fundamental asymmetry in textual survival: Sanskrit texts survived in greater numbers and in more widespread manuscript traditions than texts in other languages of Indian intellectual culture.

### 2.5 Prior work on conceptual chronology in Indian philosophy

On the dating of key texts, we draw on the following scholarly consensus positions: Witzel (1995) for the Rigveda (approximately 1500-1200 BCE); Olivelle (1998) for the early Upanishads (approximately 900-600 BCE); Gombrich (1988) and Norman (1983) for the Pali Canon (500-300 BCE); Jacobi (1884) for the Acaranga Sutra (500-400 BCE); Kalupahana (1986) and Garfield (1995) for Nagarjuna's Mulamadhyamakakarika (approximately 150-250 CE); Larson (1979) for the Samkhyakarika (approximately 350-450 CE); Mayeda (1992) for Shankara's works (approximately 788-820 CE). Where these positions are contested, we adopt more conservative date ranges and flag the uncertainty as approximate.

## 3. Temporal Attribution: Method, Sources, and Limitations

### 3.1 Design of the temporal attribution layer

We built a temporal attribution layer to provide a partial corrective. The layer consists of 40 scholarly-cited sources covering the period from approximately 1500 BCE to 1700 CE, and ~120 individual concept-datings. For each concept in the layer, we identify the earliest datable source in which the concept appears in a philosophically systematic form, and provide a scholarly citation for the dating.

Several design decisions in the layer deserve explicit justification.

**We use earliest documented attestation, not origin.** We do not claim that the earliest text we can date represents the first time a concept was thought or used. Oral traditions predate texts. Lost texts predate surviving ones. The oral composition of Vedic texts preceded their written transmission by centuries. We use the phrase earliest documented attestation deliberately: it is a claim about what the scholarly literature can currently support, not a claim about the ultimate origin of a concept. This distinction is critical. Reviewers of computational humanities papers rightly resist origin claims. Documented attestation claims are weaker, more defensible, and more honest.

**We record date ranges, not point dates.** Every source in the layer is given an early and a late date, and a certainty classification. The Rigveda is dated 1500-1200 BCE with certainty classified as approximate, following Witzel (1995). The Pali Canon is dated 500-300 BCE with certainty classified as approximate, following Gombrich (1988) and Norman (1983).

Shankara's Brahmasutra Bhashya is dated 788-820 CE with certainty classified as well-established, following Mayeda (1992). The range acknowledges that scholarly dating of ancient Indian texts is often contested, and that forcing a point estimate would falsely imply precision we do not have.

**We classify certainty at three levels.** Well-established: scholarly consensus is clear and the dating is based on multiple independent lines of evidence. Approximate: a reasonable scholarly estimate exists but there is meaningful disagreement within a century or two. Contested: substantial scholarly disagreement exists. We do not use contested datings as primary evidence for the mismatch analysis. Where datings are contested, we note the debate and flag the concept accordingly.

**The layer is a curated scholarly synthesis, not a ground truth.** The temporal dates were assigned by the authors drawing on specialist scholarship. We did not employ multiple annotators for this initial version of the layer, which means inter-annotator agreement cannot be reported. We acknowledge this as a limitation. The layer should be understood as a structured summary of current scholarly consensus on the datings of key Indian philosophical texts, not as an independent empirical measurement. Future versions of the layer should incorporate input from Indologists, Buddhist scholars, and Jain studies specialists to improve coverage and identify cases where our summaries are incomplete or contestable.

### 3.2 The three-tier dating system

Because only 24 of the 2,349 concepts in the darshana-graph have explicit scholarly citations in the temporal layer, we needed a method to assign dates to the remaining concepts in order to produce era-specific network snapshots. We developed a three-tier dating system that assigns concepts to dates at different levels of reliability, and marks each tier explicitly in all outputs.

**Tier 1: Explicit scholarly citation.** The concept has a direct entry in the temporal layer with a datable source and a scholarly citation. 24 concepts, 98 individual datings (some concepts have multiple relevant sources). These are the only dates used in the mismatch analysis (Section 4.2) and the only ones interpreted as documented attestations. They are flagged with T1 in all output tables.

**Tier 2: School proxy date.** The concept has no explicit citation in the temporal layer, but the school that mentions it most in the corpus has an established founding date in the scholarly literature. We assign the school founding date as the concept's temporal date. This is a lower bound: the concept cannot predate the school, but the concept may be considerably older. A concept discussed in Shankara's commentaries gets a proxy date of 788 CE, but Shankara was commenting on Upanishadic texts whose concepts are far older. The proxy date tells us only when the concept became visible to the specific school's textual tradition, not when the concept first appeared in Indian philosophical discourse. These concepts are flagged T2 in all output tables, and their dates are described as textual visibility boundaries rather than historical attestation dates.

**Tier 3: Unknown.** The concept belongs to a school whose founding date is not established, or appears in a context where no proxy date can be responsibly assigned. 0 concepts fell into this category in our dataset, because all schools in the darshana-graph have at least approximate founding dates in the scholarly literature.

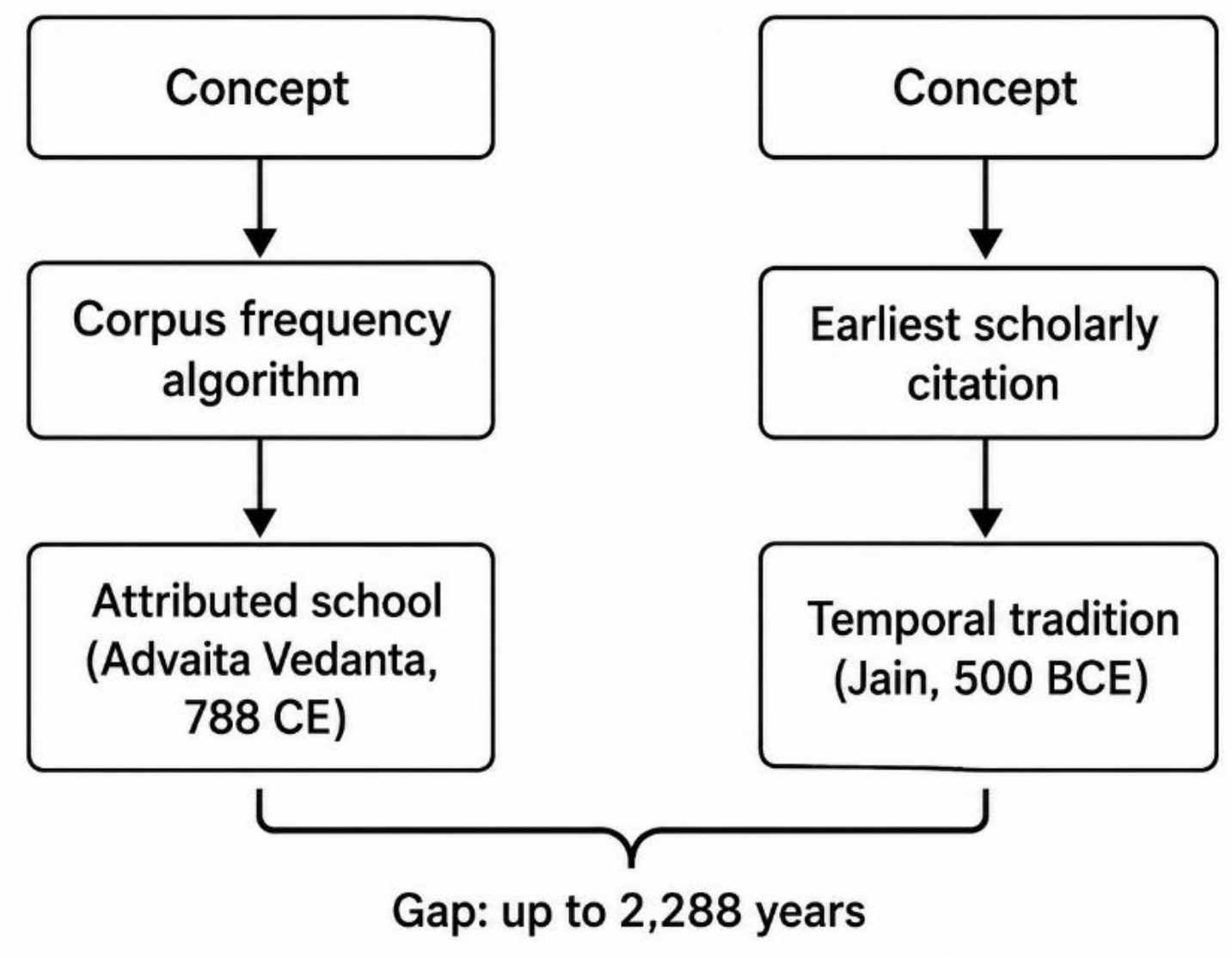

*Figure 1: Two Attribution Pipelines for Philosophical Concepts*

**3.3 The 800 CE distortion: Tier 2 proxies and corpus composition**

The most important consequence of the Tier 2 proxy system is what we call the 800 CE distortion, and it deserves full discussion before the results section because it is both a finding and a methodological limitation.

Between 300 CE and 800 CE, the era-specific network grows from 18 nodes (all Tier 1) to 1,028 nodes (24 Tier 1, 1,004 Tier 2). Of the 1,004 Tier 2 nodes that enter the network at 800 CE, 1,001 carry Advaita Vedanta proxy dates, because Advaita's school founding date of 788 CE triggers the inclusion of all concepts attributed to Advaita by corpus frequency.

This jump from 18 to 1,028 nodes does not reflect historical philosophical innovation. It reflects the following compounding factors. First, Advaita Vedanta's extensive commentarial output in the eighth through twelfth centuries produced an enormous number of concepts in the corpus. Second, Advaita's mathas provided institutional continuity for the preservation and copying of those texts. Third, Buddhist Sanskrit manuscripts from the same period were largely destroyed with the monastic universities and survive primarily in Tibetan translation, which is not included in the corpus. Fourth, Advaita's commentarial tradition absorbed and discussed concepts from Samkhya, Yoga, Mimamsa, Nyaya, and earlier Buddhist schools, meaning these older concepts appear in Advaita texts and are attributed to Advaita by corpus frequency.

The consequence is that the 800 CE era snapshot appears 97.4% Advaita Vedanta. This is not a picture of Indian philosophy at 800 CE. It is a picture of the surviving Advaita-dominated Sanskrit corpus at that date. We report this finding prominently not because it invalidates the temporal method but because it makes the corpus composition problem visible as a measurable number. The 97.4% figure is a measurement of dataset bias, not of historical reality. Making this explicit is more useful than pretending the method is unaffected by it.

The most reliable era snapshot in our analysis is 300 BCE, where all 17 nodes are Tier 1 with explicit scholarly citations. We discuss this snapshot in detail in the results section.

### 3.4 Key entries in the temporal layer: selected examples with commentary

Table 1 presents selected entries from the temporal layer chosen to illustrate the range of cases, from well-established datings to contested ones, and from cases where the temporal attribution matches the corpus-frequency attribution to cases where there is substantial divergence.

**Table 1: Selected temporal layer entries**

| Concept | Corpus attribution | Earliest attestation | Date | Certainty | Key citation | Gap |
|---|---|---|---|---|---|---|
| brahman | Advaita Vedanta | Vedic (as sacred utterance) | 1500-1200 BCE | Approximate | Witzel 1995 | 2,288 yr |
| atman | Advaita Vedanta | Vedic (as breath/self) | 1500-1200 BCE | Approximate | Olivelle 1998 | 2,288 yr |
| karma | Advaita Vedanta | Vedic/early Upanishadic | 900-600 BCE | Approximate | Olivelle 1998; Bronkhorst 2007 | 1,688 yr |
| moksha | Advaita Vedanta | Jain (Acaranga Sutra) | 500-400 BCE | Approximate | Jacobi 1884; Bronkhorst 2007 | 1,288 yr |
| maya** | Advaita Vedanta | Vedic (as magical power)** | 1500-1200 BCE | Approximate | Witzel 1995; King 1995 | 2,288 yr |
| pradhana | Advaita Vedanta | Samkhya (Samkhyakarika) | 350-450 CE | Approximate | Larson 1979 | 438 yr |
| sunyata | Madhyamaka | Madhyamaka (Nagarjuna) | 150-250 CE | Well-established | Garfield 1995 | 0 |
| ahimsa | Advaita Vedanta | Jain (Acaranga Sutra) | 500-400 BCE | Approximate | Jacobi 1884 | 1,288 year |
| alayavijnana | Yogacara | Yogacara (Vasubandhu) | 300-400 CE | Approximate | Schmithausen 1987 | 0 |
| turiya | Advaita Vedanta | Pre-Shankara Advaita | 500-700 CE | Approximate | King 1995 | 288 yr |

*** Note: Word attested in Rigveda; philosophical concept of maya as cosmic illusion enters via Gaudapada's adoption of Madhyamaka arguments (King 1995), not of Vedic origin.*

Several observations on specific entries:

**Brahman and atman.** The Rigveda uses brahman to mean sacred utterance or power, and atman to mean breath. Neither has the philosophical meaning they acquire in the Upanishads. The philosophical brahman as the absolute and atman as the individual self identical to it are developed systematically in the early Upanishads (Brihadaranyaka Upanishad 1.4.10, Chandogya Upanishad 6.8.7) at least several centuries before Shankara. Corpus-frequency attribution assigns both to Advaita Vedanta because Advaita's extensive Upanishadic commentaries mention them most often. The temporal layer identifies Vedic and early Upanishadic sources as the earliest attestation. The gap is 2,288 years.

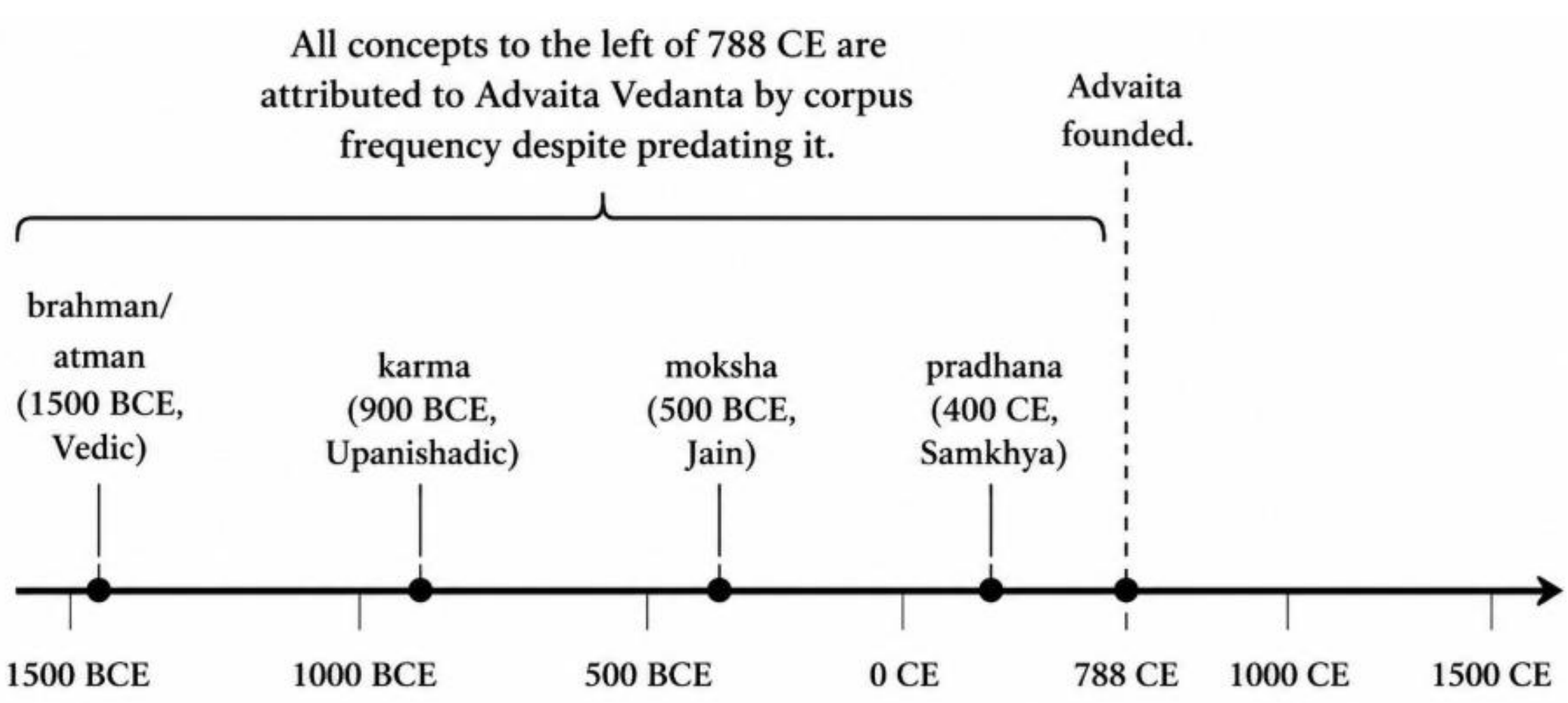

*Figure 2: Conceptual Timeline: Earliest Documented Attestation vs. Corpus Attribution*

**Moksha.** The earliest available systematic soteriological use of a term equivalent to moksha as liberation from rebirth appears in Jain sources (Acaranga Sutra, ca. 500-400 BCE, Jacobi 1884) and in early Buddhist texts predating the Upanishadic systematisation of the concept. We note that some scholars argue for a Brahmanical origin of the liberation concept. We follow Bronkhorst (2007) in treating the Shramana attestation as the earliest currently available, while acknowledging the debate. What is not contested is that the concept predates Advaita Vedanta by well over a millennium.

**Pradhana.** This is the conceptually simplest entry. Pradhana is a technical Samkhya term for primal unmanifest matter (equivalent to prakriti), documented in Ishvarakrishna's Samkhyakarika (ca. 350-450 CE, Larson 1979). Advaita Vedanta engaged extensively with Samkhya in its commentaries and discussed pradhana as part of that engagement. Corpus-frequency attribution assigns it to Advaita. The temporal layer correctly identifies it as a Samkhya concept.

**Karma.** It appears in the Rigveda as ritual action but the soteriological doctrine, i.e. karma as intentional action accumulating and determining future rebirth, is first systematically attested in Shramana traditions according to Bronkhorst (2007). The entry in Table 1 records the word's earliest textual attestation; the philosophical concept has a contested origin that the temporal layer cannot resolve.

### 3.5 Limitations of the temporal method

We have described what the temporal method attempts to do. We now describe what it cannot do and what it gets wrong, because these limitations are as important as the method itself.

The temporal layer covers 24 of 2,349 concepts with Tier 1 citations. This is 1.0% coverage. The era-specific network snapshots for the 300 BCE, 300 CE, and 800 CE periods are therefore very differently constituted. The 300 BCE snapshot has 17 Tier-1 nodes and is the most reliable. The 800 CE snapshot has 1,028 nodes, of which 1,001 are Tier 2 proxies that reflect Advaita's school founding date rather than concept-level historical attestation. Conclusions drawn from the 800 CE snapshot must be understood as conclusions about corpus composition, not about historical philosophical activity.

The Tier 2 proxy dates introduce a directional bias. By using the school founding date as the proxy date for every concept attributed to that school by corpus frequency, we preserve the same misattribution problem we are trying to correct, just at a coarser temporal resolution. A concept that Advaita absorbed from Samkhya gets an Advaita proxy date of 788 CE, which is later than its Samkhya origin but earlier than the error would suggest if we simply used corpus frequency alone. The Tier 2 system is better than no date, but it is not a solution to the attribution problem. It is a partial correction with a known directional limitation.

The temporal layer is also a synthesis of existing scholarship rather than an independent empirical measurement. Where scholarly consensus is strong, as with the dating of Shankara or the Pali Canon, the layer is relatively reliable. Where scholarship is contested, as with the Bronkhorst thesis on karma origins or the dating of some early Jain texts, the layer reflects a position in an ongoing debate. We have tried to flag these cases, but the flagging is itself a judgment call.

The temporal layer has a directional bias that should be stated explicitly. It was constructed primarily from Bronkhorst (2007), Olivelle (1998), and Witzel (1995), which are specialist sources on Brahmanical and early Upanishadic texts. Buddhist and Jain scholarship is less thoroughly represented. As a consequence, several attributions in the layer reflect word attestations in Vedic texts rather than earliest philosophical usage of the concept. Maya as cosmic illusion, avidya as root ignorance, and samsara as cyclic rebirth all have their philosophical development in Buddhist and Shramana traditions before or concurrent with Brahmanical adoption. The layer has been corrected for avidya and maya but the problem likely affects other entries. Future versions should be constructed with equal representation of Buddhist Pali Canon scholarship, Jain canonical scholarship, and Sanskrit Brahmanical scholarship.

Finally, and most fundamentally: earliest documented attestation is not origin. The oral traditions that preceded the Vedic texts, the texts that were lost in the destruction of monastic libraries, the traditions that never produced written texts: none of these are recoverable by a method that relies on datable scholarly sources. The temporal layer can correct the most obvious corpus-frequency misattributions. It cannot recover the full historical picture. No computational method can.

# 4. Results

## 4.1 Corpus-frequency attribution: top-25 betweenness centrality

Table 2 presents the top 25 concepts in the darshana-graph by betweenness centrality under corpus-frequency attribution. Betweenness centrality measures the proportion of shortest paths between all pairs of nodes that pass through a given node (Freeman 1977). A concept with high betweenness sits on many paths between other concepts and is therefore structurally central to the network: ideas travelling from one cluster to another must pass through it.

**Table 2: Top 25 concepts by betweenness centrality under corpus-frequency attribution**

| Rank | Concept | Betweenness | Corpus school | Corpus tradition |
|---|---|---|---|---|
| 1 | brahman | 0.3814 | advaita | Hindu/Advaita |
| 2 | atman | 0.2362 | advaita | Hindu/Advaita |
| 3 | krsna | 0.1143 | achintya_bhedabheda | Hindu/Vaishnava |
| 4 | krsna consciousness | 0.0606 | achintya_bhedabheda | Hindu/Vaishnava |
| 5 | karma | 0.0366 | advaita | Hindu/Advaita |
| 6 | prana | 0.0335 | advaita | Hindu/Advaita |
| 7 | knowledge | 0.0312 | advaita | Hindu/Advaita |
| 8 | isvara | 0.0289 | dvaita | Hindu/Vaishnava |
| 9 | maya** | 0.0255 | advaita | Hindu/Advaita |
| 10 | dharma | 0.0253 | advaita | Hindu/Advaita |
| 11 | self | 0.0241 | advaita | Hindu/Advaita |
| 12 | ignorance | 0.0241 | advaita | Hindu/Advaita |
| 13 | lord | 0.0238 | dvaita | Hindu/Vaishnava |
| 14 | moksha | 0.0194 | advaita | Hindu/Advaita |
| 15 | meditation | 0.0175 | advaita | Hindu/Advaita |
| 16 | soul | 0.0153 | dvaita | Hindu/Vaishnava |
| 17 | devotional service | 0.0151 | achintya_bhedabheda | Hindu/Vaishnava |

| Rank | Concept | Betweenness | Corpus school | Corpus tradition |
|---|---|---|---|---|
| 18 | vedas | 0.0131 | advaita | Hindu/Advaita |
| 19 | body | 0.0122 | advaita | Hindu/Advaita |
| 20 | light | 0.0121 | advaita | Hindu/Advaita |
| 21 | pradhana | 0.0119 | advaita | Hindu/Advaita |
| 22 | hiranyagarbha | 0.0116 | advaita | Hindu/Advaita |
| 23 | krishna | 0.0115 | achintya_bhedabheda | Hindu/Vaishnava |
| 24 | liberation | 0.0107 | advaita | Hindu/Advaita |
| 25 | yogi | 0.0106 | achintya_bhedabheda | Hindu/Vaishnava |

***Note: Word attested in Rigveda; philosophical concept of maya as cosmic illusion enters via Gaudapada's adoption of Madhyamaka arguments (King 1995), not of Vedic origin.*

The most striking feature of this table is its uniformity. Every concept in the top 25 is attributed to a Hindu school. Advaita Vedanta accounts for 16 of the 25. No Buddhist concept appears. No Jain concept appears. The highest-ranked Buddhist concept by betweenness is sunyata, which falls outside the top 25. The highest-ranked Jain concept, ahimsa, also falls outside the top 25.

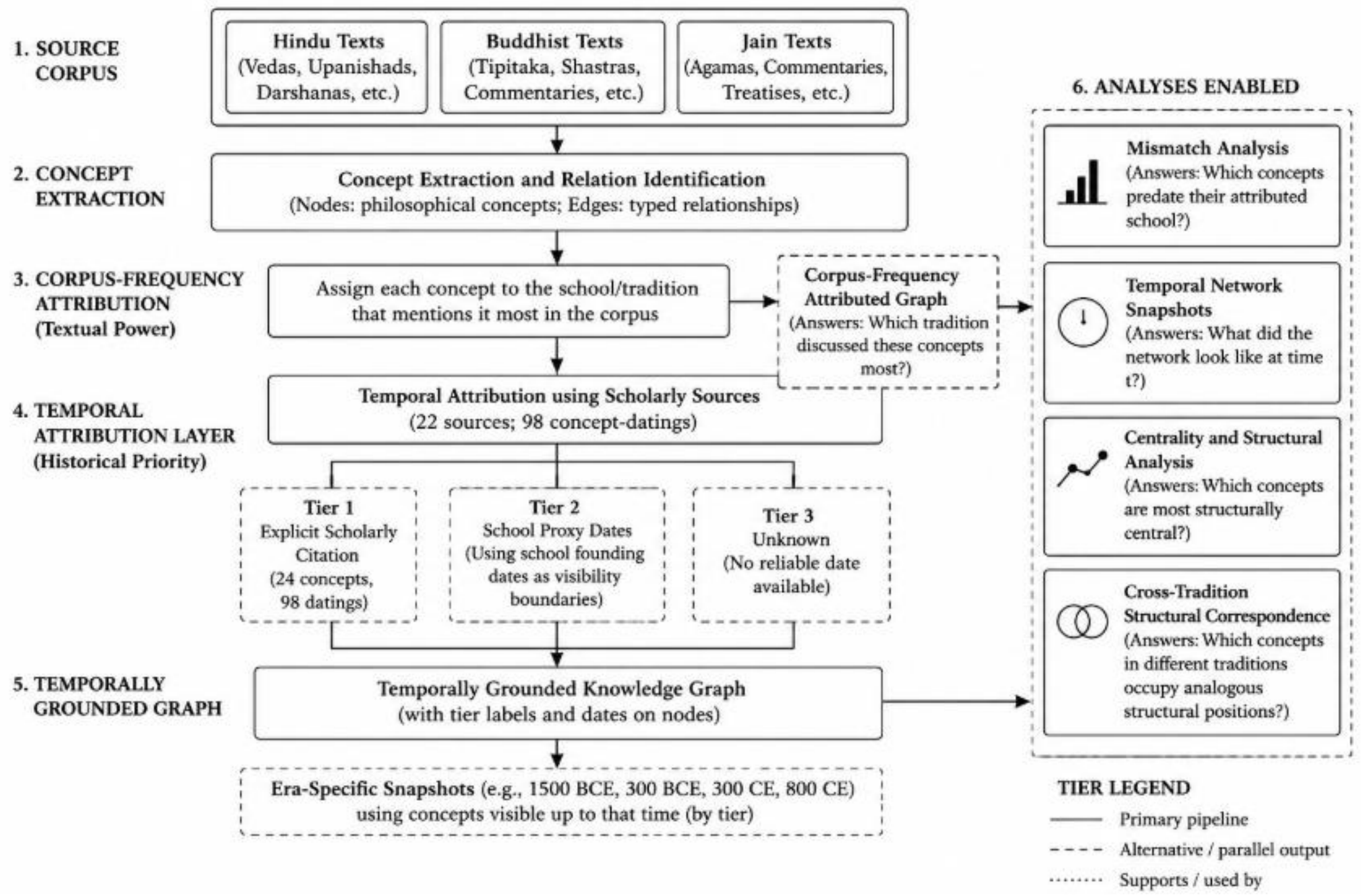

*Figure 3: Overview of the temporal attribution flow and analysis framework*

The betweenness distribution is highly skewed. Brahman at 0.38 is nearly double atman at 0.23. The top three concepts together account for a disproportionate share of total betweenness. Below rank 5 the values drop sharply and cluster around 0.01-0.03. This skew suggests a network structured around a small number of dominant hubs rather than a distributed or pluralistic structure.

We note that answering Question 1 from the measurement objectives framework, which tradition discussed these concepts most, corpus-frequency attribution is performing correctly. Advaita Vedanta did discuss these concepts most extensively in the available corpus. The problem arises when this result is interpreted as an answer to Question 2 or Question 3, which it is not designed to address.

**4.2 Temporal attribution: the mismatch analysis**

Table 3 presents the mismatch analysis for all concepts in the top 25 that have Tier-1 temporal citations in the attribution layer. A mismatch is recorded when the earliest documented attestation precedes the founding of the attributed school by more than 100 years. The 100-year threshold is chosen to provide a conservative buffer for dating uncertainty.

**Table 3: Temporal mismatch table, top-25 concepts with Tier-1 citations**

| Concept | Betweenness | Corpus school (founded) | Earliest tradition | Earliest date | Gap | Certainty |
|---|---|---|---|---|---|---|
| brahman | 0.3814 | advaita (788 CE) | Vedic | 1500-1200 BCE | 2,288 yr | Approximate |
| atman | 0.2362 | advaita (788 CE) | Vedic | 1500-1200 BCE | 2,288 yr | Approximate |
| karma* | 0.0366 | advaita (788 CE) | Vedic/early Upanishadic (word); Shramana/Buddhist (soteriological concept) | 900-600 BCE | 1,688 yr | Approximate |
| prana | 0.0335 | advaita (788 CE) | Vedic | 1200-900 BCE | 1,988 yr | Approximate |
| maya** | 0.0255 | advaita (788 CE) | Vedic (word only; philosophical concept: Buddhist-origin, see note)** | 1500-1200 BCE | 2,288 yr | Approximate |
| moksha | 0.0194 | advaita (788 CE) | Jain (Acaranga Sutra) | 500-400 BCE | 1,288 yr | Approximate |

| Concept | Betweenness | Corpus school (founded) | Earliest tradition | Earliest date | Gap | Certainty |
|---|---|---|---|---|---|---|
| pradhana | 0.0119 | advaita (788 CE) | Samkhya (Samkhyakarika) | 350-450 CE | 438 yr | Approximate |
| turiya | n/a (outside top 25) | advaita (788 CE) | Pre-Shankara Advaita | 500-700 CE | 288 yr | Approximate |
| samsara*** | n/a (outside top 25) | Sramana origin | Contested (Shramana/Buddhist, Pali Canon) | 900-600 BCE | 1688 year | Approximate |

**Note 1: Karma as a word appears in Vedas to refer to ritual sacrifices, soteriological karma doctrine originates in Shramana traditions per Bronkhorst (2007)*

*** Note 2: The word maya appears in the Rigveda; the philosophical concept enters Advaita via Gaudapada's adoption of Madhyamaka arguments (King 1995).*

****Note 3: Samsara as a term appears in the Buddhist Theravada Pali Canon and Jain texts*

Seven of the top 25 concepts have Tier-1 mismatches. The gaps range from 288 years (turiya, attributed to Advaita but documented in Gaudapada's pre-Shankara Mandukya Karika) to 2,288 years (brahman, atman, and maya, which appear in Vedic sources approximately 2,300 years before Advaita Vedanta existed as a school).

Two entries in the table, moksha and pradhana, are particularly significant for the argument about cross-tradition absorption. Moksha's earliest available systematic soteriological attestation is in Jain sources, not in any Hindu tradition. Pradhana is a technical term of Samkhya philosophy, a school that predates Advaita by several centuries. In both cases, Advaita's corpus-frequency dominance reflects extensive engagement with and commentary on these concepts in Advaita texts, not the introduction of those concepts by Advaita.

We emphasise again what this analysis can and cannot claim. The mismatch table shows that corpus-frequency attribution assigns concepts to a school that was founded centuries or millennia after those concepts are first documented in scholarly-cited sources. It does not show that Advaita Vedanta's engagement with these concepts was philosophically less important, less original, or less sophisticated than earlier traditions' engagement. Earliest documented attestation is one measure. Philosophical depth is a different measure. This analysis addresses the first and is silent on the second.

### 4.3 The betweenness centrality series across eras

Table 4 presents betweenness centrality at five key historical dates using the three-tier dating system. Each era's network includes only concepts whose assigned temporal date falls at or before that era. Tier-1 concepts (explicit citations) and Tier-2 concepts (school proxy dates) are both included, but flagged separately.

**Table 4: Betweenness centrality across five historical eras**

| Era | Network nodes | Top concept | Betweenness | Temporal tradition | Tier |
|---|---|---|---|---|---|
| 600 BCE | 10 | karma* | 0.185 | Vedic/early Upanishadic* | T1 |
| 300 BCE | 17 | brahman | 0.231 | Vedic | T1 |
| 300 CE | 18 | brahman | 0.204 | Vedic | T1 |
| 800 CE | 1,028 | brahman | 0.422 | Vedic | T1 |
| 1500 CE | 2,349 | brahman | 0.381 | Vedic | T1 |

**Note: karma at 600 BCE reflects the Upanishadic karma-rebirth doctrine; Bronkhorst (2007) argues this doctrine entered Brahmanical texts from Shramana traditions*

Several observations warrant attention.

**At 600 BCE**, only 10 concepts are present, all Tier-1 with explicit scholarly citations. The network is Vedic and early Upanishadic. Karma holds the highest betweenness (0.185). Brahman and atman are present but their betweenness scores are lower than karma's at this date (0.088 each), because their dense interconnections with the broader Advaita-era vocabulary have not yet entered the network. This is the most reliable snapshot in the table. It suggests that even in the pre-Buddhist Vedic period, karma was the structurally central concept rather than brahman.

**At 300 BCE**, the network has grown to 17 concepts. Jain and Buddhist concepts have entered: ahimsa, jiva, tapas, moksha (all Jain), and anatta, anicca, dukkha, nibbana, samadhi, sila, prajna (all Theravada Buddhist). Brahman now leads betweenness at 0.231, but the tradition composition of the network is 59% Vedic/early Upanishadic, 24% Jain, and 18% Buddhist. The network is genuinely pluralistic. This snapshot, based entirely on Tier-1 citations, is the most historically reliable picture of Indian philosophical discourse available in this analysis.

**At 300 CE**, the network adds only one new Tier-1 concept relative to 300 BCE: alayavijnana, the Yogacara Buddhist storehouse consciousness (Schmithausen 1987). The network grows by just one node. This is a significant observation: relatively few concepts in our temporal layer were introduced between 300 BCE and 300 CE, reflecting the limits of our Tier-1 coverage rather than any claim about the pace of philosophical development in this period.

**At 800 CE**, the network explodes from 18 to 1,028 nodes. We discuss this jump in detail in the following subsection.

**At 1500 CE**, the full network of 2,349 concepts is active. Brahman's betweenness has decreased from its 800 CE peak of 0.422 to 0.381, reflecting the entry of a large Vaishnava Bhakti cluster of 1,321 concepts that redistributes some betweenness away from the Advaita core. The tradition composition at 1500 CE is 56% Vaishnava, 43% Advaita, with Buddhist, Jain, Vedic, Samkhya, and Yoga concepts each representing less than 1%.

#### 4.4 The 800 CE distortion: quantifying dataset bias

The jump from 18 nodes at 300 CE to 1,028 nodes at 800 CE is the most important quantitative result in this paper, not because it tells us about Indian philosophy in 800 CE, but because it tells us about the dataset.

Of the 1,028 nodes in the 800 CE network, 24 are Tier-1 (explicit citations). The remaining 1,004 are Tier-2 proxy-dated nodes. Of those 1,004, exactly 1,001 carry Advaita Vedanta's proxy date of 788 CE. The remaining 3 Tier-2 nodes belong to Samkhya (proxy date 350 CE), Yoga (proxy date 400 CE), and Pre-Shankara Advaita (proxy date 500 CE).

The tradition composition of the 800 CE network is 97.4% Hindu/Advaita. Table 5 shows the full breakdown.

**Table 5: Tradition composition at 800 CE**

| Tradition | Nodes | Percentage |
|---|---|---|
| Hindu/Advaita (Tier 2 proxy) | 1,001 | 97.4% |
| Vedic/early Upanishadic (Tier 1) | 10 | 1.0% |
| Hindu/Samkhya (Tier 2 proxy) | 7 | 0.7% |
| Jain (Tier 1) | 4 | 0.4% |
| Theravada Buddhist (Tier 1) | 3 | 0.3% |
| Hindu/Yoga (Tier 2 proxy) | 2 | 0.2% |
| Yogacara Buddhist (Tier 1) | 1 | 0.1% |

This 97.4% figure requires careful interpretation. It is not a claim that Indian philosophy in 800 CE was 97.4% Advaita Vedanta. It is a measurement of three compounding factors.

Ryan and Ahnert (2021) have shown that network centrality metrics are robust under random archival decay. Our finding illustrates precisely the case their robustness guarantees do not cover: non-random, systemic, targeted loss. The destruction of Buddhist manuscript traditions was not random. It was concentrated, repeated, and institution-specific. The resulting dataset bias is not the kind that random-decay robustness analyses address.

Brahman's betweenness reaches its peak value of 0.422 at 800 CE. This is an increase of 0.218 from its 300 CE value of 0.204, a rise of more than 100%. The timing corresponds to Shankara's activity and to the mass entry of Advaita proxy concepts into the network. We describe this as the Advaita amplification effect: the structural centrality of brahman in the network at 800 CE is not solely a consequence of brahman's philosophical importance. It is a consequence of Advaita Vedanta's commentarial output, institutional survival, and the Tier-2 proxy mechanism simultaneously expanding the network around brahman as a hub. We deliberately use the term

Advaita amplification rather than the Shankara effect, because the data cannot distinguish between the philosophical influence of Shankara specifically, the broader expansion of the Advaita commentarial tradition after Shankara, and the corpus composition effects described above. All three are confounded in this measurement.

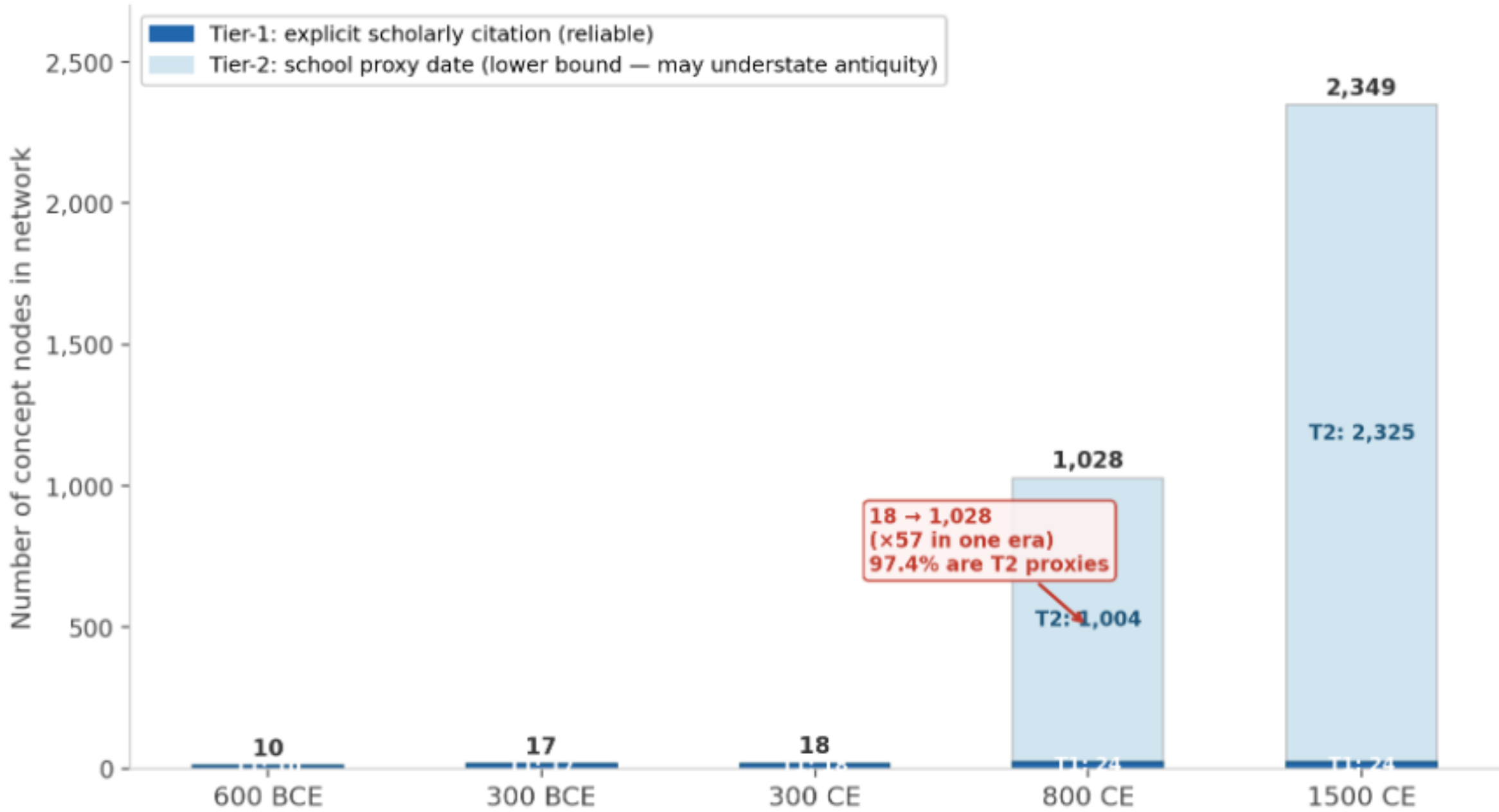

*Figure 4. Network Growth Across Eras: Tier-1 vs. Tier-2 Nodes. The jump from 18 nodes at 300 CE to 1,028 nodes at 800 CE is not historical philosophical innovation. It reflects the simultaneous entry of 1,001 Advaita Vedanta Tier-2 proxy nodes dated to 788 CE. Of the 1,028 nodes at 800 CE, 97.4% are Tier-2 proxies whose dates reflect the school's founding date rather than any concept-level attestation.*

### 4.5 The 300 BCE network: the most reliable snapshot

Because the 300 BCE snapshot relies entirely on Tier-1 concepts with explicit scholarly citations, it is the most defensible historical picture available in this analysis. Table 6 presents the full betweenness ranking at 300 BCE.

**Table 6: Full betweenness centrality at 300 BCE (17 Tier-1 nodes)**

| Rank | Concept | Betweenness | Temporal tradition | Tier |
|---|---|---|---|---|
| 1 | brahman | 0.231 | Vedic | T1 |
| 2 | karma | 0.185 | Vedic/early Upanishadic | T1 |
| 3 | atman | 0.127 | Vedic | T1 |
| 4 | jiva | 0.115 | Jain | T1 |

| Rank | Concept | Betweenness | Temporal tradition | Tier |
|---|---|---|---|---|
| 5 | samadhi | 0.000 | Theravada Buddhist | T1 |
| 6 | sila | 0.000 | Theravada Buddhist | T1 |
| 7 | prajna | 0.000 | Theravada Buddhist | T1 |
| 8 | ahimsa | 0.000 | Jain | T1 |
| 9 | tapas | 0.000 | Jain | T1 |
| 10 | moksha | 0.000 | Jain | T1 |
| 11 | samsara | 0.000 | Vedic/early Upanishadic (Contested) (appears in Theravada Buddhist Pali canon and Jain texts) | T1 |
| 12 | tat tvam asi | 0.000 | Vedic/early Upanishadic | T1 |
| 13 | sat | 0.000 | Vedic/early Upanishadic | T1 |
| 14 | prana | 0.000 | Vedic | T1 |
| 15 | maya* | 0.000 | Vedic (Word only)* | T1 |
| 16 | yajna | 0.000 | Vedic | T1 |
| 17 | soma | 0.000 | Vedic | T1 |

**Note: The word maya appears in the Rigveda; the philosophical concept enters Advaita via Gaudapada's adoption of Madhyamaka arguments (King 1995).*

At this date the network is small and sparsely connected, with only four concepts having non-zero betweenness. Brahman leads, but its betweenness (0.231) is substantially lower than its value at any later date. Jiva, the Jain concept of the individual living soul, is fourth highest, above any Buddhist concept. Moksha, which corpus-frequency attribution assigns to Advaita Vedanta, appears here as a Jain concept with a betweenness of 0.000, reflecting its presence in the network but its peripheral structural position at this date relative to brahman, karma, atman, and jiva.

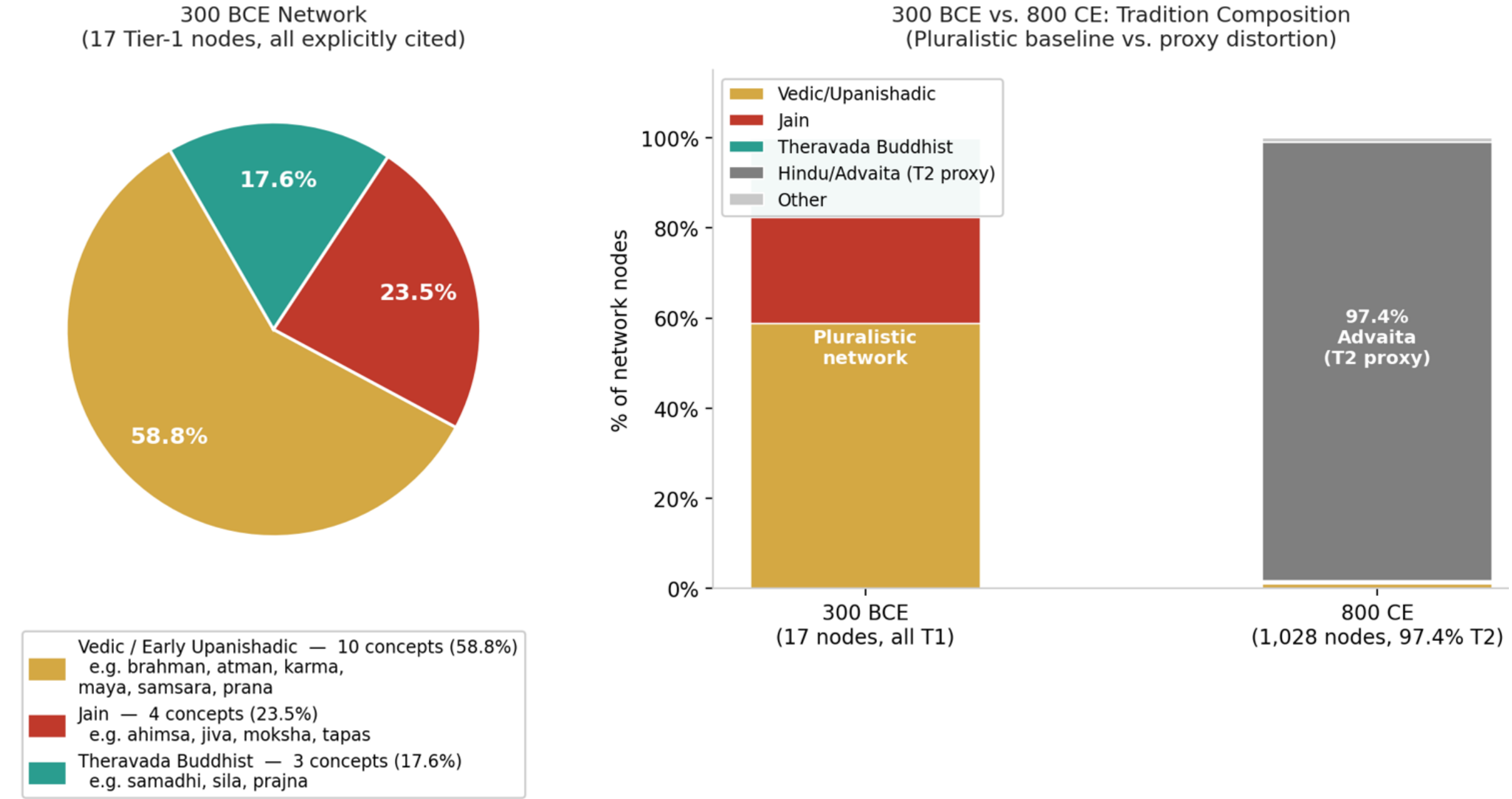

*Figure 5. Network Tradition Composition at 300 BCE (17 Tier-1 nodes) vs 800 CE*

The tradition composition is: 58.8% Vedic/early Upanishadic (10 concepts), 23.5% Jain (4 concepts), 17.6% Theravada Buddhist (3 concepts). This is not an Advaita-dominated network. It is a plural network in which Vedic, Jain, and Buddhist concepts each have significant presence.

We caution that a 17-node network is very small and the betweenness values are correspondingly sensitive to individual edges. Adding or removing a single edge can substantially change the betweenness ranking. These results should be interpreted as directional rather than precise. The value of the 300 BCE snapshot is not the specific betweenness numbers. It is the demonstration that a temporally honest, citation-grounded analysis produces a substantially different picture from corpus-frequency attribution: not a Hindu-dominated network, but a plural one.

**4.6 Cross-tradition structural correspondences**

The temporal attribution layer enables a specific class of analysis that corpus-frequency attribution makes impossible: systematic structural comparison across traditions. Because corpus-frequency attribution assigns all high-betweenness concepts to Hindu schools, no Buddhist or Jain concept receives a tradition label that can anchor a cross-tradition comparison. Temporal attribution creates the anchor set. We operationalise this as follows. For each of 48 temporally labelled concepts spanning eight traditions (Buddhist/Theravada, Buddhist/Madhyamaka, Buddhist/Mahayana, Vedic, Jain, Hindu/Samkhya, Hindu/Advaita, and Sant), we compute an ego-network feature vector consisting of: betweenness centrality, normalised degree centrality, relation-type distribution across 11 typed edges, and neighbour-tradition distribution. We then compute cosine similarity between feature vectors for all concept pairs from different traditions. This measures structural homology: similarity of network position and relational profile, not semantic or doctrinal equivalence.

We validate the method against 14 known scholarly correspondences and recover 8 of 14 (57%). Two known pairs score in the top 10: purusha-jiva (Hindu/Samkhya and Jain, sim 0.990, rank 1) and prakriti-maya (Hindu/Samkhya and Vedic, sim 0.972, rank 5). Three further known pairs score above 0.88: avidya-maya (Buddhist/Theravada and Vedic, 0.938), dukkha-samsara (Buddhist/Theravada and Vedic, 0.931), and nibbana-moksha (Buddhist/Theravada and Jain, 0.885). The recovery rate and rank positions confirm that the feature vectors are capturing genuine structural relationships rather than noise.

Beyond validation, the analysis surfaces novel structural homologies. Nibbana (Buddhist liberation) and samsara (Vedic cyclic existence) score 0.954 despite being doctrinal opposites: both function as the ultimate reference concept in their tradition's soteriology, the point against which all other concepts are understood. Moha (Buddhist delusion) and tapas (Jain austerity) score 0.959, occupying mirror-image positions: one the obstacle the path removes, the other the discipline that removes it. Cetana (Buddhist intention, the volitional factor that constitutes karma) and ajiva (Jain non-living matter, the inert substrate that jiva becomes entangled with) score 0.923, a pairing absent from the comparative philosophy literature. Both are the secondary ontological category in their tradition, the counterpart to the primary active principle. Vedana (Buddhist feeling-tone) and samsara score 0.919: both are the mechanism-of-continuation in their tradition's analysis of suffering. These pairings are structural measurements, not philosophical arguments. They identify concept pairs warranting further scholarly examination.

The analysis also produces artifacts that require acknowledgment. Soma (Vedic ritual plant deity) appears with high similarity scores in nine of the top 50 pairs. Its atypical structural position produces a feature vector that spuriously matches many peripheral concepts; high-scoring pairs involving soma should be treated with caution. Similarly, moha dominates the top 50 with nine appearances, reflecting either genuine structural centrality as an obstacle concept connected broadly throughout the Buddhist network, or a generic feature vector arising from high connectivity. Distinguishing these cases requires philological scrutiny beyond the graph's scope. The structural homology analysis is further constrained by the 1% Tier-1 temporal layer coverage. Key concepts including sunyata, anatta, and tathagatagarbha appear as nodes but have insufficient edges for reliable feature vectors, a consequence of the same corpus composition problem documented in Section 4.4. Expanding Buddhist edge density is the primary target for future work.

## 5. Discussion

### 5.1 What corpus-frequency attribution actually measures

This is a real and historically interesting quantity. The fact that Advaita Vedanta dominates the corpus-frequency attribution of the darshana-graph is itself historically informative. It tells us about the differential survival of textual traditions in Indian intellectual history, about the institutional infrastructure that supported Sanskrit Vedantic scholarship, and about the digitisation priorities of twentieth and twenty-first century Indology. These are legitimate objects of scholarly inquiry.

Collins (1998) provides the theoretical framework for understanding why this happens. His sociology of philosophies shows that intellectual networks are shaped by their material bases: the institutional structures, patronage systems, and preservation infrastructures that determine which intellectual coalitions survive and which contract. Advaita Vedanta's network of mathas, its association with royal patronage across multiple dynasties, and the continuity of Sanskrit as a cosmopolitan scholarly language (Pollock 2006) gave it the material base to produce and preserve an enormous textual output. The Buddhist universities lacked this material continuity after the twelfth century. The corpus reflects these institutional histories, and corpus-frequency attribution inherits them without acknowledgment.

Drucker (2011) makes a related point in a different register: humanistic data must be understood as capta, constructed, selected, and interpreted, rather than as neutral pre-existing facts. The darshana-graph is capta. Its edge distribution reflects human decisions about which texts to digitise, which relationships to extract, and which attribution scheme to apply. Corpus-frequency attribution naturalises these decisions as if they were neutral measurements. Temporal attribution, imperfect as it is, makes some of those decisions visible and challengeable.

### 5.2 What temporal attribution actually measures

Temporal attribution with scholarly-cited source dates measures earliest documented attestation: the earliest point in time at which a scholarly-cited source records a concept in philosophically systematic form. This is a meaningful historical quantity, and it is substantially different from corpus-frequency attribution, as the results demonstrate. Most importantly, temporal attribution cannot measure philosophical significance: the antiquity of a concept does not determine the depth of a tradition's treatment of it. The fact that brahman appears in the Rigveda as sacred utterance does not mean that the sophisticated Advaita Vedantic analysis of brahman as the non-dual absolute is philosophically less valuable than its Vedic antecedent. The temporal history of a concept and its philosophical importance are separate matters.

The measurement objectives framework introduced in Section 1.2 maps onto a three-layer distinction between what data encode (corpus composition), what algorithms compute (centrality, path lengths), and what results might mean historically or philosophically. Most concept graph papers blend these layers without distinguishing them. The implications for computational humanities follow directly: attribution method, corpus composition, and the layer of evidence being invoked should all be stated explicitly rather than assumed.

### 5.3 Structural homology as a downstream consequence of temporal attribution

The structural homology analysis reported in Section 4.6 is not an independent contribution. It is a direct consequence of temporal attribution. Under corpus-frequency attribution, every high-betweenness concept is assigned to a Hindu school, and cross-tradition structural comparison is impossible: there are no Buddhist or Jain tradition labels to serve as anchor points for comparison. The temporal attribution layer, by assigning concepts to their earliest documented tradition, creates the anchor set that makes the analysis tractable. This is the causal link between Paper 1's methodological contribution and the empirical findings: correct attribution first, then compare.

The structural homology results should be read as hypotheses, not conclusions. The method identifies concept pairs with similar structural positions in the network and presents them as candidates for scholarly examination. When the algorithm reports that nibbana and samsara are structural homologues despite being doctrinal opposites, it is not claiming these concepts are equivalent. It is identifying a structural pattern: both are the ultimate reference concept in their tradition's soteriology. Whether that structural pattern reflects genuine intellectual historical connection, parallel independent development, or algorithmic artifact requires the kind of philological and historical argument that this paper cannot provide. The contribution is identifying which pairs are worth asking. The three-tier temporal layer, the feature vector method, and the validation against known correspondences collectively provide sufficient evidence that the method is producing signal. The interpretation of that signal belongs to comparative philosophy.

### 5.4 The 800 CE distortion as a finding about datasets

The 97.4% Advaita composition of the 800 CE network is, we argue, the most important single result in this paper. It is important not as a finding about Indian philosophy but as a finding about datasets and about the Tier-2 proxy mechanism.

This is what we mean when we say that the 800 CE distortion is not a failure of the method but a finding. A method that produces visibly implausible results when its assumptions are violated is more scientifically useful than a method that produces plausible-seeming results regardless of whether its assumptions are met. The corpus-frequency method produces plausible-seeming results, of course Advaita dominates, the texts are dominated by Advaita, and this plausibility conceals the methodological problem. The Tier-2 proxy method produces a result so extreme that it forces the question: what is this actually measuring?

Future work should develop methods that distinguish between concepts that were newly introduced by a school and concepts that a school absorbed and discussed. This is a difficult annotation problem, it requires the kind of specialist historical knowledge that the temporal layer partially encodes, but it is the key missing piece for making era-specific network snapshots historically meaningful rather than corpus-composition artefacts.

### 5.5 Implications for computational humanities

The findings of this paper suggest three specific implications for computational humanities research that uses concept graphs.

**First, attribution method should be stated as a primary methodological choice, not a preprocessing detail.** The choice between corpus-frequency attribution and temporal attribution produces substantially different results. This choice should be stated, justified, and its limitations acknowledged in any paper that presents results based on concept attribution. A paper that reports betweenness centrality rankings without stating how concepts were attributed to traditions is analogous to a paper that reports regression coefficients without stating how variables were coded.

**Second, corpus composition should be reported alongside network structure.** The proportion of concepts attributed to each tradition by corpus frequency, the proportion of Tier-1 versus Tier-2 concepts in any era snapshot, and any known corpus biases (digitisation advantage, textual survival patterns, extraction pipeline artefacts) should be reported as part of the methods, not acknowledged only in limitations. A researcher reading the 800 CE network results without knowing that 97.4% of nodes are Advaita Vedanta Tier-2 proxies will draw wrong conclusions. This information should be visible, not buried.

**Third, the three-layer distinction should be made explicit.** Results should be labelled by which layer they belong to: what the data encode, what the algorithm computes, and what the results might mean. Claims at Layer 3 should be identified as interpretations rather than findings, and the gap between the computational result and the historical or philosophical claim should be stated. The phrase "the graph shows" should be distinguished from "the evidence suggests" which should be distinguished from "we interpret this as."

These three implications are minimal requirements, not aspirational ideals. They ask only that computational humanities papers state what they are measuring, acknowledge what they are not measuring, and keep the layers of evidence and interpretation distinct. This is not a critique of computational methods in the humanities. It is an argument for using those methods with the same epistemological care that distinguishes good humanistic scholarship in any medium.

Beyond correcting attribution bias, temporally grounded concept graphs enable a new class of computational humanities inquiry. When concepts are situated in historically appropriate traditions rather than attributed by corpus frequency, researchers can investigate structural correspondences across traditions without assuming doctrinal equivalence. The shortest-path analysis demonstrates this possibility concretely: the Buddhist concept sunyata and the Advaita concept brahman are separated by a single conceptual bridge, maya, a distance shorter than many concepts within the same tradition. The Jain concept moksha and the Buddhist nibbana occupy the same structural role as terminal soteriological goals in their respective networks. These are not claims of philosophical identity. They are claims of structural homology: different traditions use different concepts to occupy the same functional position in the network of philosophical thought. Computational concept graphs, once temporally grounded, make these homologies measurable for the first time. This may eventually support a form of computational translation between philosophical vocabularies, not asserting that one tradition's terms mean the same as another's, but identifying which concepts perform analogous structural work across traditions. That is a research programme this paper opens rather than completes.

### 5.6 Broader applicability

The measurement problem identified in this paper is not specific to Indian philosophy or to the darshana-graph. It arises wherever a computational knowledge graph is built from a corpus in which textual production, survival, and digitisation are systematically uneven across the traditions represented.

In all these cases, a computational concept graph built without temporal sourcing will reflect textual power rather than intellectual history. Making this explicit, in the methods, in the results, and in the interpretation, is a minimal requirement for responsible computational humanities scholarship. The tools we propose here, particularly the distinction between textual power, historical priority, and philosophical significance, and the three-tier uncertainty system for concept dating, are transferable to any of these contexts. They do not solve the problem of uneven textual survival. No computational method can. But they make the problem visible, measurable, and explicitly acknowledged, which is the necessary first step toward addressing it.

## 6. Conclusion

### 6.1 Summary of findings

This paper has examined a specific and consequential methodological problem in computational knowledge graphs built from historical philosophical texts: the conflation of corpus-frequency attribution with historical priority or philosophical significance.

We demonstrated this problem using the darshana-graph, a knowledge graph of 28,322 documented philosophical relationships across Hindu, Buddhist, and Jain traditions. Under corpus-frequency attribution, the top 25 concepts by betweenness centrality are all attributed to Hindu schools, with Advaita Vedanta dominating. Under temporal attribution using 40 scholarly-cited sources and ~120 concept-datings, seven of those top 25 concepts have earliest documented attestations that predate the school to which corpus-frequency assigns them by between 288 and 2,288 years. The concept moksha, attributed to Advaita Vedanta by corpus frequency, has its earliest available systematic soteriological attestation in Jain sources more than 1,200 years before Advaita Vedanta existed. The concept pradhana, similarly attributed to Advaita, is a technical term of Samkhya philosophy documented centuries before Shankara.

We also identified and quantified a distortion within the temporal method itself. Between 300 CE and 800 CE the era-specific network grows from 18 to 1,028 nodes, with 97.4% of new nodes carrying Advaita Vedanta proxy dates derived from the school's founding date of 788 CE rather than concept-level historical attestation. This growth reflects corpus composition, the scale of Advaita's commentarial output, the destruction of Buddhist manuscript traditions, and the digitisation advantage of Sanskrit texts, rather than historical philosophical activity at that date. We report this distortion explicitly because it is the clearest demonstration in the paper of the broader principle: computational results about historical texts are simultaneously shaped by intellectual history, textual survival, institutional power, and dataset construction choices, and conflating these produces systematically misleading conclusions.

The most historically reliable snapshot available in this analysis is the 300 BCE network, based on 17 Tier-1 concepts with explicit scholarly citations. At that date, the network is 59% Vedic and early Upanishadic, 24% Jain, and 18% Buddhist. It is not an Advaita-dominated network. It is a genuinely pluralistic network in which the three major Shramana and Brahmanical traditions have comparable structural presence. This stands in stark contrast to the full corpus-frequency graph, which presents Indian philosophical thought as overwhelmingly Hindu.

### 6.2 The methodological contribution

The core methodological contribution of this paper is the measurement objectives framework introduced in Section 1.2 and applied throughout: the distinction between three questions that concept graphs can be asked to answer, each requiring a different method and producing a different kind of answer.

Corpus-frequency attribution answers the question of which tradition discussed a concept most in the available corpus. It is the appropriate method when textual engagement is the research object. It is a misleading method when historical priority or philosophical significance is the research object.

Temporal attribution with scholarly-cited source dates answers the question of which tradition first documented a concept in a datable source. It is a partial corrective for the most obvious corpus-frequency misattributions. Its own limitations, 1% Tier-1 coverage, directional bias in Tier-2 proxy dates, reliance on contested scholarly datings, are as important to state as the method itself.

Neither method answers the question of which tradition developed the most philosophically sophisticated or original treatment of a concept. This question requires historical scholarship, philosophical argument, and textual analysis that no computational method can substitute for.

### 6.3 Future directions

Future work should expand Tier-1 coverage collaboratively with Indologists, Buddhist scholars, and Jain studies specialists; build tradition-specific graphs to avoid cross-tradition corpus composition problems; develop temporal graph visualisations; run community detection with temporal constraints; and apply the measurement objectives framework to other historical philosophical corpora including medieval European scholasticism, Chinese classical philosophy, and Islamic philosophy.

This paper provides the attribution-corrected graph that makes two further investigations possible. The first concerns temporal transmission: tracing the historical sequence by which concepts moved between traditions through documented figures, which the transmission layer begins to model. The second concerns structural correspondence: identifying which concepts across traditions occupy analogous network positions, not as a claim of doctrinal identity but as a computable measure of structural homology. Combining temporal grounding, semantic embeddings, and role-based similarity measures may eventually support a form of computational cross-tradition concept alignment, identifying the closest structural analogue of a Madhyamaka concept within Advaita, or a Jain soteriological concept within Buddhism, as hypotheses for scholarly examination rather than settled claims. The darshana-graph, once temporally attributed, is the infrastructure this programme requires.

### 6.4 Closing statement

The darshana-graph is capta, not data: every edge it contains reflects decisions about which texts to digitise, which relationships to extract, and which attribution scheme to apply. Corpus-frequency attribution measures textual power. Temporal attribution measures earliest documented attestation. Neither measures philosophical significance. Stating which of these is being measured, and acknowledging what is not being measured, is the minimum requirement for computational humanities scholarship that can be trusted. Correctly attributed, the darshana-graph suggests that the philosophical traditions of India are not separate islands of thought but different vocabularies for navigating a shared conceptual space, and for the first time, we can begin to measure the bridges between them.